\documentstyle[pra,aps,manuscript]{revtex}

\input{epsf}

\begin{document}
\title{Realization of an Optimally Distinguishable Multi-photon Quantum
Superposition.}
\author{Francesco De Martini and Fabio Sciarrino}
\address{Dipartimento di Fisica and Istituto Nazionale di Fisica della\\
Materia, Universit\'{a} ''La Sapienza'', Roma 00185, Italy}
\maketitle

\begin{abstract}
We report the successful generation of an {\it entangled} multiparticle
quantum superposition of pure photon states. They result from a multiple 
{\it universal} cloning of a single photon qubit by a high gain,
quantum-injected parametric amplifier. The {\it information preserving}
property of the process suggests for these states the name of ''{\it %
multi-particle qubits''}.{\it \ }They are ideal objects for investigating
the emergence of the classical world in quantum systems with increasing
complexity, the decoherence processes and may allow the practical
implementation of the universal 2-qubit logic gates.
\end{abstract}

\pacs{PACS numbers: \tt\string}

\narrowtext

Since the golden years of quantum mechanics the interference of classically
distinguishable quantum states, first epitomized by the famous ''{\it %
Schroedinger-Cat}'' apologue \cite{Schr35} has been the object of extensive
theoretical studies and recognized as a major conceptual paradigm of Physics 
\cite{Cald83,Legg85}. In modern times the sciences of quantum information
(QI)\ and quantum computation deal precisely with collective processes
involving a multiplicity of interfering states, generally mutually entangled
and rapidly de-phased by decoherence \cite{Zure81}. In many respects the
implementation of this intriguing classical-quantum condition represents
today an unsolved problem in spite of recent successful studies carried out
mostly with atoms and ions \cite{Brun96,Monr96}. The present work reports on
a nearly decoherence-free all optical scheme based on the quantum-injected
optical parametric amplification (QI-OPA) of a single photon in a quantum
superposition state of polarization $(\pi )$, i.e. a $\pi -encoded$ {\it %
qubit }\cite{DeMa98,DeMa01}.

Conceptually, the method consists of transferring the well accessible
condition of quantum superposition characterizing a single-photon qubit, $%
N=1 $, to a {\it mesoscopic}, i.e. multi-photons amplified state $M>>1$,
here referred to as a ''{\it multi-particle qubit}'' ({\it M-qubit}). In
quantum optics this can be done by injecting in the QI-OPA the single-photon 
{\it qubit, }$\alpha \left| H\right\rangle +\beta \left| V\right\rangle $,
here expressed in terms of two mutually orthogonal $\pi -$states, e.g.
horizontal and vertical linear $\pi ^{\prime }s$: $\left| H\right\rangle $, $%
\left| V\right\rangle $. In virtue of the general \ {\it information
preserving} \ property of the OPA,\ the generated multi-particle state is
found to keep the {\it same} superposition character and the interfering
capabilities of the injected qubit, thus realizing the most relevant and
striking property of \ the {\it M-qubit} condition \cite{DeMa01}. Since the
present scheme basically realizes the {\it deterministic} $1\rightarrow M$ 
{\it universal} {\it optimal quantum} {\it cloning} {\it machine} (UOQCM),
i.e. able to copy {\it optimally} any unknown input qubit into $M>>1$ copies
with the same fidelity, the output state will be necessarily affected by 
{\it squeezed-vacuum (SV)} noise. This one partially spoils the {\it exact},
i.e. ''classical'' distinguishability of the two interfering terms of the
superposition. This is but a further manifestation of \ the quantum\
''no-cloning theorem'' within the present \ {\it optimal} process. In a more
technical perspective, since any UOQCM can be designed to redistribute {\it %
optimally }the initial information into many output channels, the present
scheme is expected to find useful QI\ applications, e.g. in quantum
cryptography \cite{Bech99}, \cite{Galv00} and in error correction schemes 
\cite{Brub01}. In addition, a similar method was successfully adopted
recently to realize the first $1\rightarrow 2$ UOQCM and the first universal
quantum NOT-gate \cite{DeMa02,Pell03,Lama02}.

Let's refer to the apparatus: Fig. 1. The OPA active element was a nonlinear
(NL)\ crystal slab (BBO:\ $\beta -$barium borate), 1.5 mm thick cut for Type
II\ phase-matching, able to generate by spontaneous parametric down
conversion (SPDC) $\overrightarrow{\pi }-entangled$ $\ $pairs of photons.
Precisely, the OPA {\it intrinsic phase} was set as to generate by SPDC {\it %
singlet} entangled states on the output modes, a condition assuring the {\it %
universality} of the cloning transformation by the QI-OPA scheme \cite
{Lama02,DeMa98,DeMa02}. The photons of each pair were emitted with equal
wavelengths (wls) $\lambda =795nm$ over two spatial modes $-{\bf k}_{1}$ and 
$-{\bf k}_{2}$ owing to a SPDC process excited by a coherent pump UV\ field
provided by a Ti:Sa Coherent MIRA mode-locked pulsed laser coupled to a
Second Harmonic Generator (SHG) and associated with a mode with wavevector
(wv) $-{\bf k}_{p}$ and wl $\lambda _{p}=397.5nm$. The average UV\ power was 
$0.25W$, the {\it pulse repetition rate }${\it (}rep$-$r{\it )}$, $%
r_{UV}=7.6\ast 10^{7}$ $s^{-1}$, and the {\it coherence time} of each UV
pulse as well of the generated single photon pulses were $\tau _{coh}=140$
fs. The SPDC process implied a 3-wave NL parametric interaction taking place
towards the right hand side (r.h.s.)\ of Fig. 1. The UV\ pump was
back-reflected over the mode ${\bf k}_{p}\ $onto the NL\ crystal by a
spherical mirror ${\bf M}_{p}$, with $\mu $-metrically adjustable position 
{\bf Z}, thus exciting the main OPA ''cloning''\ process, towards the left
hand side (l.h.s.) of Fig.1. By the combined effect of two adjustable
optical UV waveplates (wp) $(\lambda /2$ + $\lambda /4)$ acting on the
projections of the linear polarization ${\bf \pi }_{p}$ of the UV field on
the fixed optical axis of the BBO crystal for the $-{\bf k}_{p}$ and ${\bf k}%
_{p}$ counter propagating \ excitation processes, the SPDC\ excitation was
always kept at a very low level while the main OPA amplification could reach 
{\it any large intensity, }as we shall see shortly. Precisely, by smartly
unbalancing the orientation angles $\vartheta _{%
{\frac12}%
}$ and $\vartheta _{%
{\frac14}%
\text{ }}$of the UV $wp^{^{\prime }}s$, the probability ratio of SPDC\
emission towards the r.h.s. of Fig.1, of two simultaneous correlated photon
pairs and of a single pair was always kept below $3\times 10^{-2}$ in any 
{\it High Gain} condition. One of the photons of the SPDC\ emitted pair,
back-reflected by a fixed mirror ${\bf M}$, was {\it re-injected} onto the
amplifying NL\ crystal by the input mode ${\bf k}_{1}$, while the other
photon emitted over mode $(-{\bf k}_{2})\ $excited the detector $D_{T}$, the 
{\it trigger }of the overall {\it conditional }experiment. The detectors $%
(D)\;$were single-photon SPCM-AQR14. A proper setting\ of {\bf Z} secured
the space-time overlapping into the NL crystal of the interacting\
re-injected pulses with wl's $\lambda _{p}$ and $\lambda $, and then
determined the optimal QI-OPA condition. The time optical walk-off effects
due to the crystal birefringence were compensated by inserting in the modes $%
{\bf k}_{1}$, ${\bf k}_{2}$ and $-{\bf k}_{2}$ three fixed X-cut quartz
plates $Q$ and one $\lambda /4\ $wp. Before re-injection into the NL\
crystal, the {\it pure} input qubit on mode ${\bf k}_{1}$, $\left| \Psi
\right\rangle _{in}$= $(\widetilde{\alpha }\left| \Psi \right\rangle 
{\alpha  \atop in}%
+\widetilde{\beta }\left| \Psi \right\rangle 
{\beta  \atop in}%
)$, $\left| \widetilde{\alpha }\right| ^{2}+\left| \widetilde{\beta }\right|
^{2}=1$, represented by the Bloch sphere shown in Fig. 1, underwent unitary $%
SU(2)$ {\it rotation} $\Phi \ $transformations: $\widehat{U_{i}}\equiv \exp
(-i\sigma _{i}\Phi /2)\ $around the three Cartesian axes $i=x,y,z\ $by the
combined action of the $\lambda /2$ wp $WP_{T}$, of the adjustable {\it %
Babinet Compensator} $B$ and of the polarizing beam-splitter $PBS_{T}$
acting on mode $(-{\bf k}_{2})$ in virtue of the nonlocality correlating the
modes $-{\bf k}_{1}$ and $-{\bf k}_{2}$. These $SU(2)$ transformations are
represented in Fig.1 (inset) by circles drawn on the surface of a
Bloch-sphere. The main OPA process, i.e. acting towards the l.h.s. of Fig.1
on the injected qubit $\left| \Psi \right\rangle _{in},\;$was characterized
by two different excitation regimes, establishing two corresponding {\it %
sizes} of the output {\it M-qubit}:

A)\ \ {\it Low Gain} $(LG)$\ regime, characterized by a low excitation UV\
energy $(3.5nJ)$ per pulse, leading to a small value of the NL\ parametric ''%
{\it gain}'' :\ $g=0.07.$

B)\ \ {\it High Gain}\ $(HG)$\ regime, characterized by a larger value (by a
factor $\approx 16$) of the {\it gain}: $g=1.13$.\ This condition was
attained by a further amplification of the UV ''pump'' beam by a Ti-Sa
regenerative amplifier Coherent-REGA\ operating at a pulse $rep$-$r$: $%
r_{UV}=2.5\ast 10^{5}$ $s^{-1}$, with pulse duration: $180fs$:\ Fig.1.

Let us re-write the state $\left| \Psi \right\rangle _{in}$ by expressing
the interfering states as Fock product states: $\left| \Psi \right\rangle 
{\alpha  \atop in}%
$= $\left| 1\right\rangle _{1h}\left| 0\right\rangle _{1v}\left|
0\right\rangle _{2h}\left| 0\right\rangle _{2v}\equiv \left|
1,0,0,0\right\rangle $; $\left| \Psi \right\rangle 
{\beta  \atop in}%
$= $\left| 0,1,0,0\right\rangle $, accounting respectively for 1 photon with
horizontal $(h)$ polarization on the input ${\bf k}_{1}$, vacuum state on
the input ${\bf k}_{2}$, and 1 photon with vertical $(v)$ $\pi $ on ${\bf k}%
_{1}$, vacuum state on the mode ${\bf k}_{2}$. The solution of the QI-OPA
dynamical equations is found to be expressed by the {\it M-qubit}: $\left|
\Psi \right\rangle \equiv (\widetilde{\alpha }\left| \Psi \right\rangle
^{\alpha }+\widetilde{\beta }\left| \Psi \right\rangle ^{\beta })$, with 
\begin{equation}
\left| \Psi \right\rangle ^{\alpha }\equiv \gamma \sum\limits_{i,\
j=0}^{\infty }(-\Gamma )^{i}\Gamma ^{j}\sqrt{i+1}\ \left|
i+1,j,j,i\right\rangle ;\ \ \left| \Psi \right\rangle ^{\beta }\equiv \gamma
\sum\limits_{i,\ j=0}^{\infty }(-\Gamma )^{i}\Gamma ^{j}\sqrt{j+1\ }\left|
i,j+1,j,i\right\rangle
\end{equation}
where: $\gamma \equiv C^{-3}$, $C\equiv \cosh g$ \ and $\ \Gamma \equiv
\tanh g$ \cite{DeMa02}. These interfering {\it entangled}, multi-particle
states are {\it ortho-normal}, i.e. $\left| ^{i}\left\langle \Psi \right|
\left| \Psi \right\rangle ^{j}\right| ^{2}$= $\delta _{ij}$ $\left\{
i,j=\alpha ,\beta \right\} $\ and {\it pure, }i.e. fully represented by the
operators: $\rho ^{\alpha }=(\left| \Psi \right\rangle \left\langle \Psi
\right| )^{\alpha }$, $\rho ^{\beta }=(\left| \Psi \right\rangle
\left\langle \Psi \right| )^{\beta }$.\ Hence the {\it pure} state $\left|
\Psi \right\rangle $ is an entangled\ quantum superposition of two
multi-photon {\it pure} {\it states }and bears the {\it same} superposition
properties of the injected single-photon qubit. The genuine quantum
signature of this superposition, namely the {\it non-definite positive}
character of the Wigner function $W(\alpha ,\beta )$\ of the output $\left|
\Psi \right\rangle $ in the 8-dimensional complex phase-space $\alpha _{j}$, 
$\beta _{j}$, $j=1,2$, was confirmed by a previous theoretical analysis \cite
{DeMa98}. For the sake of completeness, consider the overall output density
operator $\rho \equiv \left| \Psi \right\rangle \left\langle \Psi \right| \ $%
and his {\it mixed-state} reductions over the $\overrightarrow{\pi }-vector$
spaces relative to the modes ${\bf k}_{1}$and ${\bf k}_{2}$: $\ \rho
_{1}=Tr_{2}\rho \ $;$\ \rho _{2}=Tr_{1}\rho $. These ones may be expanded as
a weighted superpositions of $p-square$ matrices of order $p=(n+2)$, the
relative weight $\Gamma ^{2}$ of each two successive matrices being
determined by the parametric {\it gain}. Note that $\Gamma ^{2}$ approaches
asymptotically the unit value for large $g$.\ In turn, the $p-square$
matrices may be expressed as sum of $2\times 2$ matrices as shown by the
following expressions: 
\begin{equation}
\rho _{1}=\gamma ^{2}\sum\limits_{n=0}^{\infty }\Gamma ^{2n}\times
\sum\limits_{i=\ 0}^{n}\left[ 
\begin{array}{cc}
\left| \widetilde{\beta }\right| ^{2}(n-i+1) & \widetilde{\alpha }^{\ast }%
\widetilde{\beta }\sqrt{(i+1)(n-i+1)} \\ 
\widetilde{\alpha }\widetilde{\beta }^{\ast }\sqrt{(i+1)(n-i+1)} & \left| 
\widetilde{\alpha }\right| ^{2}(i+1)
\end{array}
\right]
\end{equation}
written in terms of the Fock basis: $\left\{ \left| i\right\rangle
_{1h}\left| n-i+1\right\rangle _{1v}\text{;\ }\left| i+1\right\rangle
_{1h}\left| n-i\right\rangle _{1v}\right\} $. Correspondingly: 
\begin{equation}
\rho _{2}=\gamma ^{2}\sum\limits_{n=0}^{\infty }\Gamma ^{2n}\times
\sum\limits_{i=0}^{n+1}\left[ 
\begin{array}{cc}
\left| \widetilde{\beta }\right| ^{2}(n-i+1) & -\widetilde{\alpha }^{\ast }%
\widetilde{\beta }\sqrt{(n-i+1)i} \\ 
-\widetilde{\alpha }\widetilde{\beta }^{\ast }\sqrt{(n-i+1)i} & \left| 
\widetilde{\alpha }\right| ^{2}i
\end{array}
\right]
\end{equation}
in terms of the Fock basis:$\left\{ \left| n-i\right\rangle _{2h}\left|
i\right\rangle _{2v}\text{;\ }\left| n-i+1\right\rangle _{2h}\left|
i-1\right\rangle _{2v}\right\} $. Interestingly, the value $n$ appearing in
Eqs. 2, 3 coincides with the number of photon pairs generated by the QI-OPA\
amplification. Note the {\it non-diagonal }character{\it \ }of these
matrices\ implying the quantum superposition property of the overall state.
Furthermore the interfering states possess the {\it nonlocal} property of
all entangled quantum systems. It is worth noting that the Von Neumann
entropies $S(\rho _{j})=Tr(\rho _{j}\log _{2}\rho _{j})$, $j=1,2$ are equal,
thus implying the same {\it degree of mixedeness }on the 2 output channels%
{\it :} $S(\rho _{1})=$ $S(\rho _{2})$ \cite{DeMa02,Benn96}.

The actual $\left| \Psi \right\rangle $ experimental detection of the output
states could be undertaken by measurements taken either (a) on the
injection, i.e. cloning mode ${\bf k}_{1}$, or (b) on the {\it anticloning}
mode ${\bf k}_{2}$, or (c) on both output modes. The option (b) was selected
since there the field on mode ${\bf k}_{2}$ is not affected by the input
qubit in absence of the QI-OPA\ action and then any registered {\it %
interference} effect is by itself an unambiguous demonstration of the {\it %
M-qubit} condition. This condition was carefully verified experimentally. In
addition, it was checked that only {\it non interfering} SV-noise was
detected on the output mode ${\bf k}_{2}$, in absence of the injection qubit 
$\left| \Psi \right\rangle _{in}$.

Since the $1^{st}-$order interference property of any quantum object is
generally expressed by the $1^{st}$-$order$ {\it correlation-function} $%
G^{(1)}$ \cite{Wall94}, this quantity was measured by two detectors $D_{2}\ $%
and $D_{2^{\ast }}$ coupled to the mutually orthogonal fields $\widehat{c}%
_{j}(t)$,$\ j=1,2$\ emerging from the polarizing beam-splitter $PBS_{2}$
inserted in the output ${\bf k}_{2}$: $G%
{(1) \atop 2j}%
\equiv \left\langle \Psi \right| \widehat{N}_{j}(t)\left| \Psi \right\rangle 
$. In the present context the $G^{(1)}$ are expressed by ensemble averages
of the number operators $\widehat{N}_{j}(t)\equiv \widehat{c}_{j}^{\dagger
}(t)~\widehat{c}_{j}(t)$, written in terms of the detected fields: $\widehat{%
c}_{1}(t)\equiv 2^{-%
{\frac12}%
}[\widehat{b}_{h}(t)+\widehat{b}_{v}(t)]$ , $\widehat{c}_{2}(t)\equiv 2^{-%
{\frac12}%
}[\widehat{b}_{h}(t)-\widehat{b}_{v}(t)]$. There the $\widehat{b}_{i}(t)$, $%
i=h,v$, defined on the QI-OPA\ output ${\bf k}_{2}$, underwent a $45^{0}\pi
- $rotation by the $\lambda /2-$wp $WP_{2}$ before injection into $PBS_{2}$:
Fig. 1. The same operations transformed the basis $\{h,v\}$ into $\{H,V\}$.
By expressing the injected qubit as: $\left| \Psi \right\rangle _{in}\equiv
\left( \alpha \left| \Psi \right\rangle _{in}^{\alpha }+\beta e^{i\varphi
}\left| \Psi \right\rangle _{in}^{\beta }\right) $ with $\alpha $ and $\beta 
$ real numbers, the $G%
{(1) \atop 2j}%
$ could be given as: $G%
{(1) \atop 2H}%
=\overline{n}+%
{\frac12}%
\overline{n}\left[ 1+2\alpha \beta \cos \varphi \right] $; $G%
{(1) \atop 2V}%
=\overline{n}+%
{\frac12}%
\overline{n}\left[ 1-2\alpha \beta \cos \varphi \right] $, showing the
superposition character of the output average field per mode with respect to 
$\alpha ,\beta $ and $\varphi $, where $\overline{n}\equiv \sinh ^{2}g$. By
comparison of this result with the corresponding {\it non interfering}
averages $G%
{(1) \atop 21,vac}%
=G%
{(1) \atop 22,vac}%
=\overline{n}$ taken in absence of the injected qubit, i.e. over the {\it %
input} {\it vacuum state,} the signal-to-noise ratio for the {\it M-qubit}
detection: was found:$\ S/N$ $=$ $2,$ for $\varphi =0$ and $\alpha =\beta
=2^{-%
{\frac12}%
}$.

The above result immediately suggests, as a demonstration of the $1^{st}$-$%
order$\ ${\bf \pi -}$interference of the {\it M-qubit}, to draw the{\it \
fringe} patterns expressing the quantity: $\Delta G%
{(1) \atop 2}%
\equiv (G%
{(1) \atop 2H}%
-G%
{(1) \atop 2V}%
)$ and bearing the fringe ''visibility'': ${\cal V}=(G%
{(1) \atop 2\max }%
-G%
{(1) \atop 2\min }%
)/(G%
{(1) \atop 2\max }%
+G%
{(1) \atop 2\min }%
)=2\alpha \beta /3$. The experimental values of $G^{(1)}$ were obtained by
coincidence measurements involving the couples of detectors: $[D_{2}D_{T}]$
and $[D%
{\ast  \atop 2}%
D_{T}]$. Note that the fringe patterns, reported in Fig. 2 and the
corresponding values of ${\cal V}$ were strongly affected by the
superposition character of the injection qubit and by the corresponding $%
SU(2)$ transformations expressed by the closed paths drawn on the Bloch
sphere: Fig.1 (inset). In order to show more clearly the quantum efficiency
of the {\it M-qubit} for the two ''gain'' regimes, the {\it relative
coincidence rates,} $\xi =r_{SC}/r_{UV}$ were reported in Fig. 2, where $%
r_{SC}$ expressed the values of $\Delta G%
{(1) \atop 2}%
\equiv (G%
{(1) \atop 2H}%
-G%
{(1) \atop 2V}%
)$. The experimental best-fit curves drawn in Figure 2 were found to
reproduce very closely the quantum theoretical results.

In the $LG$ excitation condition the {\it average} number of
quantum-interfering photon pairs was $\overline{N}\approx 0.009$ while the $%
HG$ condition it was $\overline{N}\approx 4$, corresponding to the
realization of a much ''fatter'' {\it M-qubit}. The value of {\it average}
number of generated pairs was $\overline{N}=3\overline{n}$ in virtue of the
stimulated emission process. They were estimated by accounting for the
overall {\it quantum efficiency }of the $D$ detectors : $QE\approx 18\%$.
Calibrated attenuation filters placed in front of the $D^{\prime }s$ assured
the condition of single-photon detection in the $HG$ regime. By the Eqs. 2,
3 a straightforward evaluation of the output photon-pair distribution $p(n)$
could be carried out. By this one, the probability of generating a number of
pairs $N\geq 2\overline{N}=8$, i.e. a quantum superposition in excess of $16$
photons, was found to be $P(\overline{N})\equiv \sum\limits_{k=2\overline{N}%
}^{\infty }p(k)\approx 14\%$. By moving the injected state with $\alpha
=\beta $ over the Bloch sphere, a visibility$\ V_{th}=33\%$ of the
fringe-pattern was expected from theory. Due to experimental imperfections,
mostly due to the mode selection before detection, we have measured $%
V\approx 4\%$ $(4\%)$ with a 2-coincidence detection measurement scheme and $%
V\approx 25\%$ $(7\%)$ with a 4-coincidence scheme respectively for the $LG$
and $HG$ regimes. The results above are expected to be {\it linearly scaled}
by adoption of a more efficient NL crystal and of a more powerful UV source.

Unlike other systems involving atoms or ions \cite{Monr96,Brun96} a
limitation of our method consists of the impossibility of controlling the
''distance'', $d$ on the {\it phase-space} of the interfering multi-particle
states by means of the QI-OPA parameters. This condition, implied by the
results of the Wigner function analysis \cite{DeMa98} can be expressed here
by the Hilbert-Schmidt $(d)$ of the interfering states: $d(\rho ^{\alpha
};\rho ^{\beta })=Tr\left[ (\rho ^{\alpha }-\rho ^{\beta })^{2}\right] $ 
\cite{Nie00}. This quantity, not affected by the amplification process, is
expressed in our case by: $d(\rho ^{\alpha };\rho ^{\beta })_{in}$= $d(\rho
^{\alpha };\rho ^{\beta })_{out}$= $2$. However, as a lucky counterpart of
this condition, the de-coherence of our system can be deemed generally
irrelevant as it is mostly determined by the stray reflection losses on the
surfaces of the optical components of the apparatus. Accordingly, a number
of photons in the range $(10^{2}\div 10^{3})$ could be made simultaneously
excited in quantum superposition without sizable decoherence effects: this
is but one of the commendable qualities of the device. Of course, the
injection into the OPA of a many particle state will results in a better $%
S/N $ ratio. The M-qubit condition attained by injection into OPA\ \ of a
symmetrized 2 photon state is presently in progress in our Laboratory.

In summary, we have realized the interference of classically distinguishable 
{\it multi-particle}, {\it orthonormal}, {\it pure} quantum states. These
characteristics indeed express the peculiar properties of the system
originally proposed by Schroedinger \cite{Schr35} with a relevant additional
feature: the {\it entanglement} of the interfering states. The adoption of
the OPA to {\it clone} {\it universally} a single qubit in a large gain
regime witnesses the scalability of the information preserving property of
this device\ in the quantum domain. Furthermore and very important, unlike
other systems involving atoms, e.g. in hardly accessible optical traps or
superconducting cavities \cite{Monr96,Brun96}, the {\it M-qubit} generated
by our system is directly accessible to measurement or to further
exploitation, e.g. by injection into a QI {\it gate}. \ In a more conceptual
perspective, the present realization could open a new trend of studies on
the persistence of the validity of several crucial laws of quantum mechanics
for entangled mixed-state systems of increasing complexity \cite{Brun96} and
on the violation of Bell-type inequalities in the multi-particles regime 
\cite{Reid02}. This work has been supported by the FET European Network on
QI and Communication (Contract IST-2000-29681: ATESIT), by INFM (PRA\
''CLON'')\ and by MIUR (COFIN 2002). We thank V. Buzek, S. Popescu, for
stimulating discussions and D. Pelliccia for early experimental
collaboration.

\centerline{\bf Figure Captions}

\vskip 8mm

\parindent=0pt

\parskip=3mm

Figure 1. Layout of the {\it quantum-injected OPA }apparatus. The device 
{\bf A} represents the Regenerative Ti:Sa laser Amplifier to attain the High
Gain (HG)\ dynamical condition. INSET: Bloch sphere representation of the $%
SU(2)$ unitary transformations applied to the input qubit.

Figure 2. Multi-photon interference fringe patterns expressed by the {\it %
relative coincidence rate,} $\xi =\eta (r_{SC}/r_{UV})$ proportional to the
ratio of the 2-detector coincidence rates expressing the quantity: $\Delta G%
{(1) \atop 2}%
\equiv (G%
{(1) \atop 2H}%
-G%
{(1) \atop 2V}%
)$ and of the repetition rate of the UV pump excitation in $LG$ and $HG$
regimes $(\eta =$ $7.6\times 10^{7})$. The patterns correspond to the $SU(2)$
transformations shown in Fig.1 (inset) and affecting the injected qubit. $%
G^{(1)}$ are the $1^{st}-order$ {\it correlation functions}.

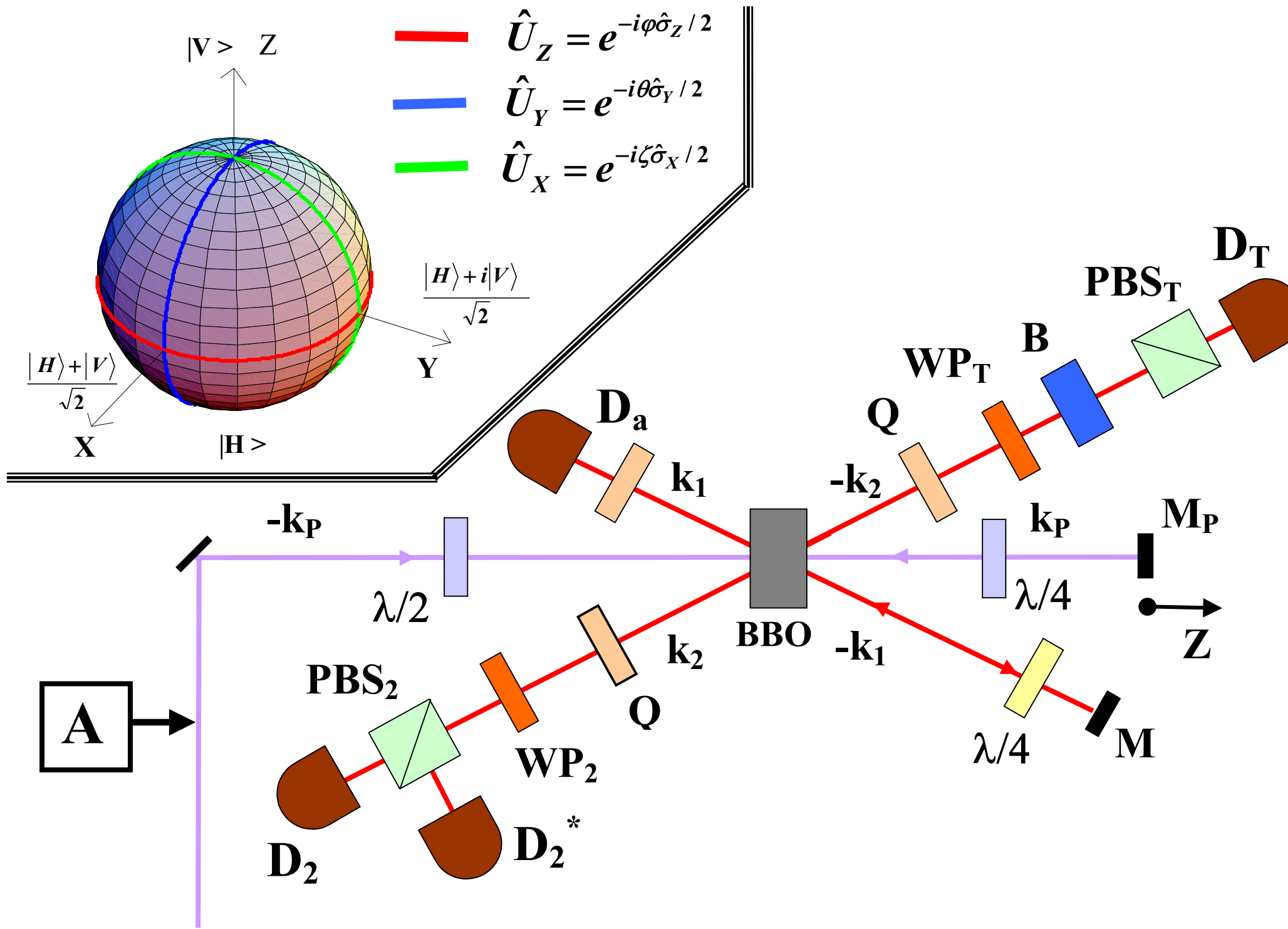
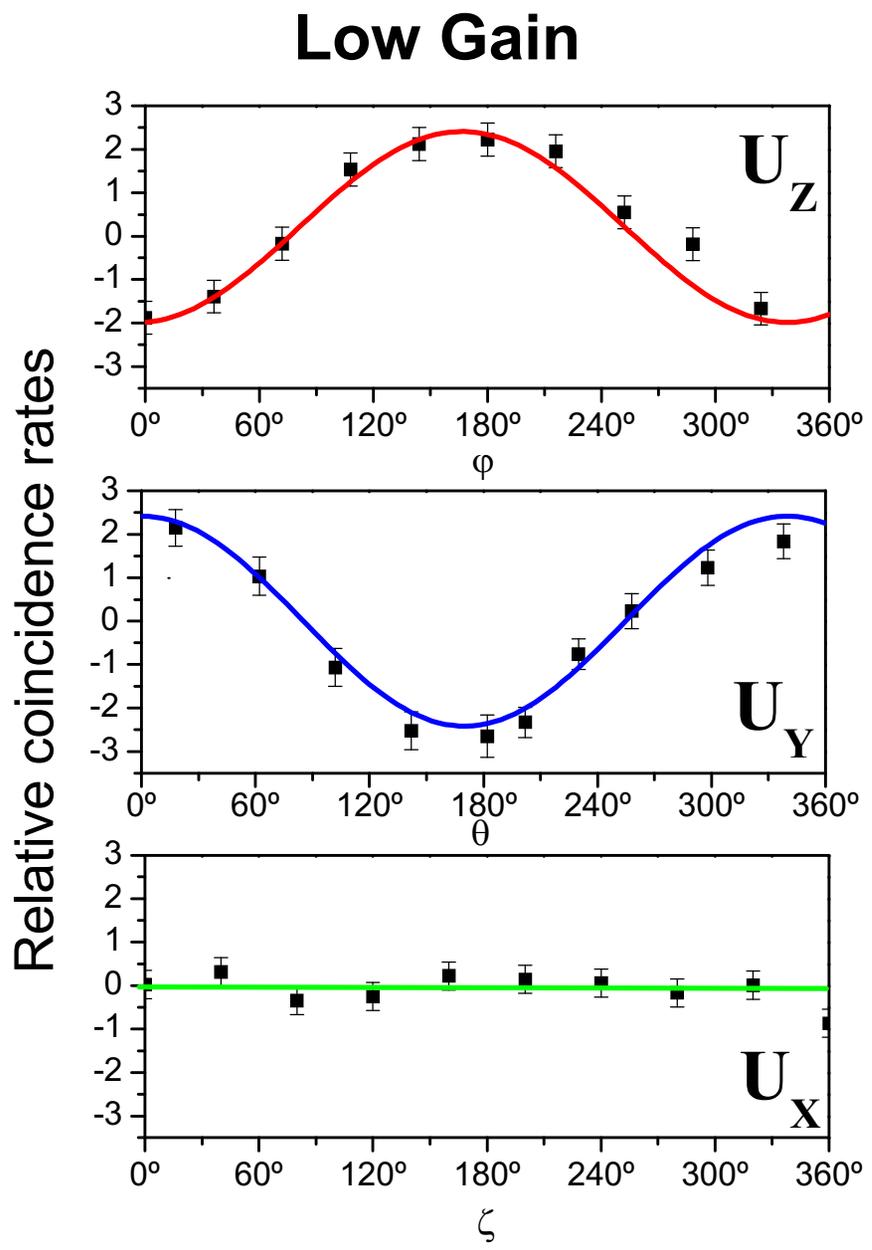
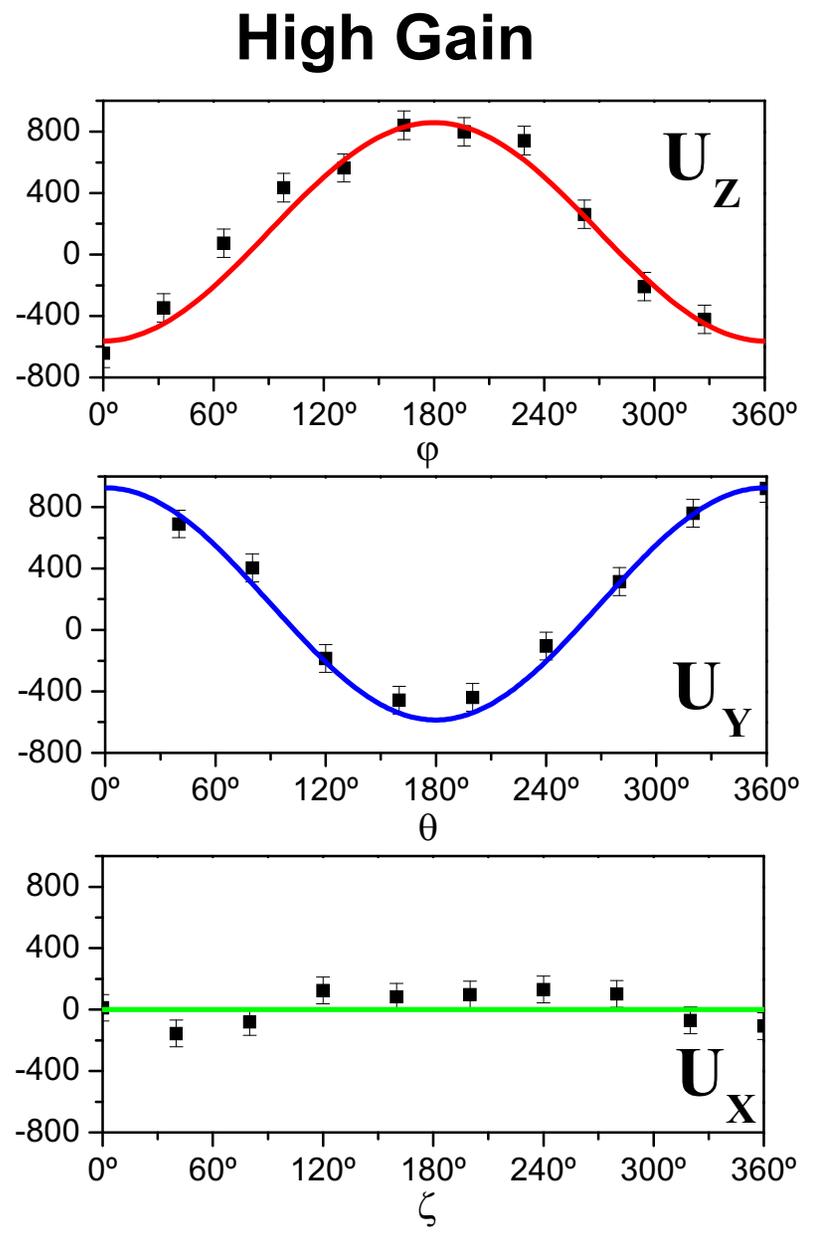
\end{document}